# A Quantum Dot Plot Generation Algorithm for Pairwise Sequence Alignment

Author(s): Joseph B. Clapis

May 2021

**Document No: MTR200363**
**Bedford, MA**

# Abstract

The Quantum Pairwise Sequence Alignment (QPSA) algorithm offers exponential speedups in data alignment tasks. However, it relies on an open problem of efficiently encoding the classical data being aligned into quantum superposition. Previous works suggest the use of specialized nonlinear Kerr media to implement a black-box oracle that achieves the superposition. We provide an alternative, explicit construction of this oracle called the Quantum Dot Plot (QDP), which is compatible with universal gate machines. We evaluate QDP's operational complexity via analysis of the quantum machine instructions generated by the Q# and Qiskit software frameworks for various sample circuits. Our analysis confirms that the data encoding scheme is exponentially difficult, precluding QDP's (and thus, QPSA's) employment on general-purpose quantum computers. Nevertheless, we find utility in estimating the circuit depth and run time of both QDP and QPSA for IBM's superconducting architecture and AQT's trapped ion architecture for direct comparison and overall practicality purposes.

# Table of Contents

# List of Figures

# List of Tables

# 1 Introduction

In the last few years, there have been many publications of novel quantum algorithms that offer computational speedups in a wide variety of domains. These speedups range from quadratic to exponential. In some cases, they address problems that can still be solved with conventional computers today, but may be expensive or time-consuming to do. In other cases, they solve problems that are completely impractical to execute on classical hardware.

Our work studies the viability of implementing these algorithms on universal gate-based quantum computers for practical use cases. We achieve this by exploring the decomposition of such algorithms into well-defined quantum circuits (e.g. no unknown subroutines), implementing the circuits using quantum software frameworks, and performing resource counts on the assembly code produced for specific platforms by quantum compilers. The intent of this research is to identify gaps in algorithm representations that preclude their implementation, close those gaps if possible, and understand the physical hardware capabilities required to leverage the algorithm for practical applications.

By studying the assembly code generated by quantum software, this process explores the instructions that will be run on the machine itself. It provides high fidelity counts for resource requirements such as circuit width (number of qubits required to execute), circuit depth (number of gates in the critical, or largest serial, path), and direct count of each type of native gate that the device supports. The metrics, in turn, can be used to inform approximate time frames of when the algorithm could potentially be used practically, based on the evolutionary and maturity trends of various universal quantum computer architectures.

In this paper, we apply our exploration process to the Quantum Pairwise Sequence Aligner (QPSA) algorithm [1]. QPSA is a quantum algorithm designed to find a near-optimal alignment of two data sequences. "Alignment" refers to the process of adjusting the two sequences such that they have the largest number of matching elements between them. For example, consider the following two binary data sequences R and Q:

```
R = 0011010010011110
Q = 1011001101001001
```

The ideal alignment here is to simply shift Q four elements to the left:

```
R' = ----0011010010011110
Q' = 1011001101001001----
```

The alignment process is allowed to add or delete individual elements to the sequences as necessary, introduce gaps, and move or invert subregions. Each of these operations (shifting, adding, deleting, moving, and inverting) is typically assigned a cost. QPSA finds the set of adjustments that provides the best alignment with the lowest cost. This is a heuristic algorithm, so it isn't guaranteed to find the *optimal* result, but it does find one within some specified tolerances for number of matches and cost.

Many classical sequence alignment algorithms already exist, and are employed in various use cases today. The most common use case for them is to align a well-known *reference sequence* (typically labeled **R**), which is not modified, with an unknown *query sequence* (labelled **Q**) which can be modified. This allows a direct comparison between the two that is helpful in identifying what the unknown sequence is, matching it to a set of candidates, or gaining some other kind of insight into it. The sequences themselves can be any arbitrary form of data, so this process has found usage in natural language processing (to determine what word was just spoken or match a voice sample to a known individual), bioinformatics (to identify an unknown sample of DNA or determine how closely one sample matches another), and other domains.

QPSA belongs to this family of algorithms, but it uses a quantum computer to provide an exponential speedup in the alignment process. It is a typical 3-stage algorithm that involves some classical preprocessing of the data, a quantum subroutine with measurements, and subsequent classical postprocessing of the measured results. Similar to Shor's algorithm, this process usually repeats several times and each iteration further refines the alignment until it finally reaches an optimal result. The details of the classical pre- and postprocessing steps are beyond the scope of this report; instead, we will focus exclusively on the quantum subroutine.

## 1.1 The Quantum Pattern Recognition Subroutine

The quantum subroutine in QPSA is called the Quantum Pattern Recognition (QPR) algorithm [2]. QPR is a relatively simple algorithm that operates on a binary sequence of data and finds periodicity, or patterns, within that sequence. It uses the Quantum Fourier Transform (QFT), or more specifically, the *inverse* QFT to highlight these patterns. The overall quantum circuit diagram for QPR is shown in Figure 1.

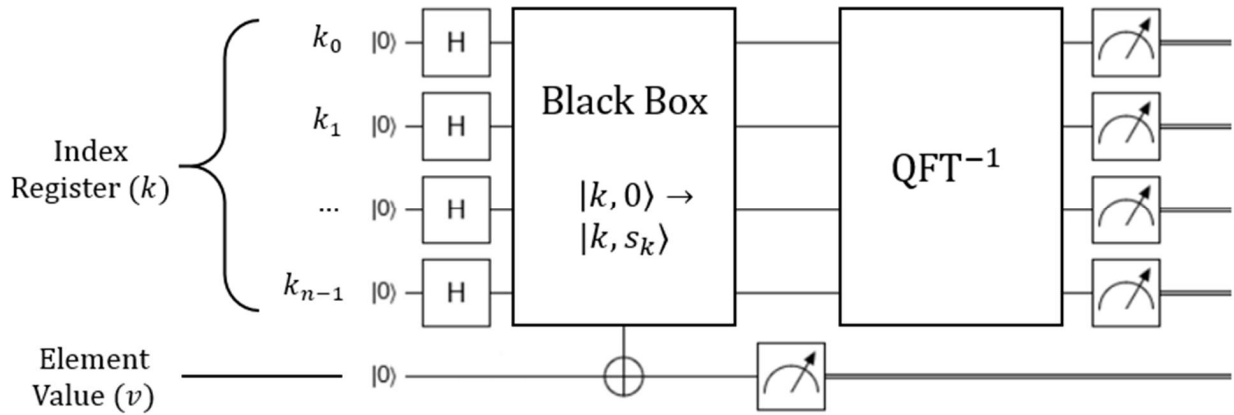

**Figure 1. The circuit diagram for QPR.**

The index register $|k\rangle$ represents the index of an element in the sequence, so it must be large enough to capture the index of the final element. It is usually assumed that the sequence's length, $N$, is a power of two ($N = 2^n$) which makes it simple to put $|k\rangle$ into a uniform superposition via the Hadamard gate. If $N$ is not a power of two, the sequence can just be padded with zeros until it is a power of two (described in section 2.2.1). The element value qubit $|v\rangle$ represents a value of one of the sequence's elements – since QPR is restricted to binary sequences, only a single qubit is needed to represent these values.

The Black Box is a subroutine that entangles $|k\rangle$ and $|v\rangle$ so that for any value of $|k\rangle$, $|v\rangle$ will always contain the bit of data in the sequence at index $|k\rangle$. In other words, the Black Box encodes the entire sequence into a quantum "lookup table" where $|k\rangle$ is the key (the index of the element) and $|v\rangle$ is the value (the value of the element at that index). More formally, the circuit starts with all qubits in the $|0\rangle$ state:

$$|k, v\rangle = |0,0\rangle \tag{1}$$

Applying the Hadamard gate to all of the qubits in $|k\rangle$ puts it into a uniform superposition of all possible values:

$$|k, v\rangle = \frac{1}{\sqrt{N}} \sum_{i=0}^{N-1} |i, 0\rangle \tag{2}$$

Finally, the Black Box encodes the sequence into the entanglement between $|k\rangle$ and $|v\rangle$:

$$|k, v\rangle = \frac{1}{\sqrt{N}} \sum_{i=0}^{N-1} |i, s_i\rangle \tag{3}$$

where $s_i$ corresponds to the $i$-th element of the sequence.

After the Black Box, QPR requires a mid-circuit measurement of $v$. This removes the entanglement between $k$ and $v$ and partially collapses $k$. If $v$ is measured as $|0\rangle$, then the only terms remaining in $k$ will be the indices where the sequence contained the value 0. Any indices where the sequence contained the value 1 will be removed from the superposition. Likewise, if $v$ is measured as $|1\rangle$, then $k$ will only contain the indices where the sequence contained the value 1. After this measurement, the inverse QFT is performed on $k$ which will amplify any trends or patterns in the remaining indices. Finally, all of the qubits of $k$ are measured to recover some information about these trends. This process is repeated multiple times until enough pattern data is retrieved that the user can make a meaningful decision from it.

## 1.2 Dot Plots

To leverage QPR, QPSA constructs a 2D graph from the two data sequences known as a *dot plot* [3]. Dot plots place the index of one sequence on one axis, and the index of the second sequence on the other. For each point on the graph, if the two sequences have the same element at those two indices, a black dot is placed at that coordinate. Dot plots effectively encode the differences and similarities between the two sequences into a 1-bit-per-pixel image. They are widely used in the bioinformatics industry to visually inspect and align relatively small sequences (with a few hundred elements) [1]. An example of a dot plot is shown in Figure 2.

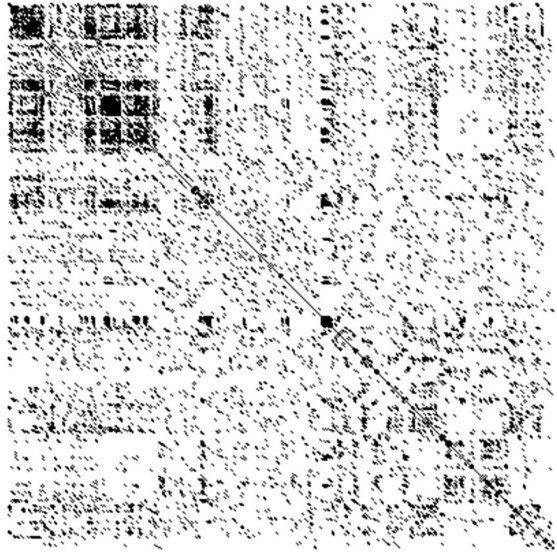

**Figure 2. Dot plot of a human zinc-finger transcription factor (GenBank NM_002383) against itself to show self-similarity.**[1]

QPSA requires the construction of a dot plot for the arbitrary reference and query sequences, $S_R$ and $S_Q$, which is then passed as input into QPR.

Let us say that $S_R$ will be represented by the $x$ axis, and its length is a power of 2 for convenience. If its length is $W = 2^w$, then the image will be $W$ pixels long. Similarly, if $S_Q$ is represented by the $y$ axis and its length is $H = 2^h$, then the image will be $H$ pixels tall. In this way, the index register $|k\rangle$ of QPR represents the 2D coordinates $(x, y)$ of a pixel in the dot plot via the following formula:

$$k = y * W + x \qquad (4)$$

Thus, the register $|k\rangle$ can be thought to represents two smaller registers $|x\rangle$ and $|y\rangle$ with lengths $w$ and $h$, respectively:

$$|k, v\rangle = |y, x, v\rangle$$

Here $y$ is placed before $x$ because in the typical algorithmic approach to processing $x$ and $y$ as a set of nested loops, $y$ is treated as the outer loop and $x$ is treated as the inner loop. Combining this interpretation with the formula for the dot plot's data, we can rephrase equation (3) to arrive at the following state vector:

$$|y, x, v\rangle = \frac{1}{\sqrt{WH}} \sum_{i=0}^{H-1} \sum_{j=0}^{W-1} |i, j, v_{ij}\rangle, \qquad v_{ij} = \begin{cases} 0, & S_{R,i} \neq S_{Q,j} \\ 1, & S_{R,i} = S_{Q,j} \end{cases} \qquad (5)$$

Thus, when used in the context of QPSA, the QPR Black Box encodes the entire dot plot into the registers $x, y,$ and $v$ via entanglement between them.

---

[1] Photo by user Opabinia regalis of Wikimedia, available under a Creative Commons Attribution-Share Alike 3.0 Unported license. Available at https://commons.wikimedia.org/wiki/File:Zinc-finger-dot-plot.png.

## 1.3 Quantum Speedup

Whereas the original dot plot occupied $2^w * 2^h * 1$ bits, the Black Box representation only occupies $w + h + 1$ qubits. For example, if both sequences were ~1,000,000 entries long and thus required 20 bits each for their indices, the classical dot plot would occupy $2^{20} * 2^{20} = 128$ GB of classical memory, but the QPR version would only occupy $20 + 20 + 1 = 41$ qubits. It follows that applying the QFT on the $|x\rangle$ and $|y\rangle$ registers requires exponentially fewer gates than the classical version, and thus QPSA provides an exponential decrease in both memory and runtime requirements.

Note that this does not take error correction or noise compensation into account, which is necessary for current universal quantum computing hardware and would increase the number of qubits required by some orders of magnitude (but still provide an exponential speedup over the classical dot plot).

## 1.4 Original Black Box Implementation via Non-Linear Kerr Devices

The reference implementation of the Black Box subroutine put forth by the QPR and QPSA papers relies on the existence of a specialized, purpose-built optical device that utilizes non-linear Kerr media. The device uses an entangled photon to represent the $|v\rangle$ qubit, and it has two sets of refractors made of a material that exhibits the Kerr effect. One set (the *horizontal* set) contains $w$ refractors, each of which is controlled by a corresponding qubit in $|x\rangle$; the other set (the *vertical* set) contains $h$ refractors and is controlled by the qubits of $|y\rangle$. When the control qubits for each horizontal refractor are in the $|1\rangle$ state, the refractors deflect the path of incoming photons horizontally by an angle that is carefully calibrated to match the index of their control qubits. The same applies for the vertical set. On the opposite side of the device is a photon sensor that is covered by a physical mask which represents the complete dot plot image. After passing the refractors, the $|v\rangle$ photon will be deflected in such a way that it arrives at the sensor in position $(x, y)$. If the sensor detects it, then the mask did not block it and thus $|v\rangle$ is 0. If the sensor does not detect it, the mask blocked it and thus $|v\rangle$ is 1.

There are several problems with this implementation. First, and most importantly for our purposes, such a device is not applicable to the general-purpose universal gate computers being developed today by the industry. The requirement on specialized hardware precludes QPSA from being executable on such machines. Second, this implementation requires the classical generation of the dot plot in order to create the sensor mask, thus greatly reducing its overall efficiency. Third, there are a number of papers that claim that Kerr media may have difficulties providing such functionality in single-photon systems [4] [5] [6].

# 2 Quantum Dot Plot Generation

In this section we propose the Quantum Dot Plot (QDP) algorithm, which conforms to the Black Box specification and is comprised of the Hadamard, Pauli-X, CNOT, and CCNOT (Toffoli) unitary gates. This composition is compatible with any universal quantum computer since, by definition, such computers can synthesize any unitary gate from their native gate sets.

QDP entails the following high-level steps:

1. Prepare the necessary registers.
2. Encode $S_R$ and $S_Q$ into quantum registers with the entanglement-based key-value representation required by QPR.
3. Compare the values of $S_R$ and $S_Q$ to generate the dot plot.

Once QDP is finished, QPR can be executed on a subset of the registers that it requires. Those registers can then be measured, which will provide the information required by QPSA.

To describe the algorithm in detail, we will first specify the associated variables:

- Let $S_R$ be the reference sequence.
- Let $S_Q$ be the query sequence.
- Let $W$ be the length of $S_R$, and let $w$ be the number of bits required to represent $W$. To trivialize superposition preparation, $W$ should be a power of two ($W = 2^w$) large enough to capture all of the indices. If the actual width is not a power of two, the sequence can be padded (discussed later) to inflate it to a power of two without issue.
- Let $H$ be the length of $S_Q$, and let $h$ be the number of bits required to represent it, following the same size and padding rules as $W$.
- Let $d$ be the number of bits needed to represent the largest data element of either sequence.

For example, consider the case where we want to align two sequences of DNA, each of which has 256 base pairs (elements). Both $W$ and $H$ are 256, which means $w$ and $h$ are 8. There are four possible base pairs (Adenine, Cytosine, Guanine, and Thymine), thus $d$ only needs to be 2.

## 2.1 Register Initialization

QDP requires five registers to operate:

- $|x\rangle$, of size $w$, represents the index of an element in $S_R$ (the x-coordinate on the dot plot).
- $|D_R\rangle$, of size $d$, represents the element of $S_R$ at index $x$.
- $|y\rangle$, of size $h$, represents the index of an element in $S_Q$ (the y-coordinate on the dot plot).
- $|D_Q\rangle$, of size $d$, represents the element of $S_Q$ at index $y$.
- $|v\rangle$ is a single-qubit register that represents the value of the dot plot at coordinates $(x, y)$. Since the dot plot is a 1-bit-per-pixel image, $v$ only needs to have one qubit.

To begin, all five registers must be initialized to $|0\rangle$:

$$|x, D_R, y, D_Q, v\rangle = |0,0,0,0,0\rangle \tag{6}$$

Next, $x$ and $y$ are put into a full uniform superposition with the Hadamard gate:

$$|y\rangle = \frac{1}{\sqrt{2^h}} \sum_{i=0}^{2^h-1} |i\rangle \tag{7}$$

$$|x\rangle = \frac{1}{\sqrt{2^w}} \sum_{j=0}^{2^w-1} |j\rangle \qquad (8)$$

The registers are now initialized and ready for the next step.

## 2.2 Sequence Encoding

The goal of this step is to encode the original sequences into qubits. To maintain the exponential reduction in storage requirements, we represent $S_R$ using the registers $x$ and $D_R$, and $S_Q$ in the registers $y$ and $D_Q$, similar to QPR's required representation. More formally, we need to achieve the following states:

$$|x, D_R\rangle = \frac{1}{\sqrt{2^w}} \sum_{i=0}^{2^w-1} |i, S_{R,i}\rangle \qquad (9)$$

$$|y, D_Q\rangle = \frac{1}{\sqrt{2^h}} \sum_{j=0}^{2^h-1} |j, S_{Q,j}\rangle \qquad (10)$$

To achieve this, we elected to leverage the encoding scheme used in the Novel Enhanced Quantum Representation of images (NEQR) algorithm [7]. NEQR itself is simply a name that refers to this particular paradigm of encoding an array of classical information in the entanglement between an index register and a data register, such as the ones presented in (9) and (10). We are using it for a somewhat unorthodox purpose here; rather than being used to encode a 2D image, we are simply encoding a raw 1D array of arbitrary data[2]. However, the encoding system fundamentally achieves the same result.

The NEQR encoding scheme approaches the problem of encoding classical data into qubits from the classical Programmable Logic Array (PLA) perspective. Essentially, it creates a unique logic circuit for the given array that takes an index as input and produces the data at that index as output. This can be done from a "brute-force" perspective by simply creating a logic sequence for each index, and then OR-ing them all together.

As an example, consider the simple binary sequence in (11) where $W = 8$, $w = 3$, and $d = 2$:

$$Ex = [00,01,11,10,01,10,11,00] \qquad (11)$$

The brute force approach would generate this logic table (using the most-significant bit [MSB] convention):

| $Index_2$ | $Index_1$ | $Index_0$ | $Data_1$ | $Data_0$ |
|---|---|---|---|---|
| 0 | 0 | 0 | 0 | 0 |
| 0 | 0 | 1 | 0 | 1 |

---

[2] In our initial experiments, we attempted to generate the dot plot classically and encode the entire image via NEQR, but this proved to produce circuits that were computationally prohibitively expensive. Because the data produced was of size $w * h$, the required circuit depth was $O(w * h)$, whereas QDP as described here only requires a circuit depth of $O(w + h)$. For $w = h$, QDP offers a quadratic speedup over simply encoding the entire dot plot via NEQR: $O(2n)$ vs. $O(n^2)$.

| 0 | 1 | 0 | 1 | 1 |
|---|---|---|---|---|
| 0 | 1 | 1 | 1 | 0 |
| 1 | 0 | 0 | 0 | 1 |
| 1 | 0 | 1 | 1 | 0 |
| 1 | 1 | 0 | 1 | 1 |
| 1 | 1 | 1 | 0 | 0 |

**Table 1. Brute-force PLA synthesis for the example in (11).**

To encode this data in qubits, this approach would create a new multicontrolled NOT gate for each row. The index register $x$ would be the controls, and the relevant data qubits in $D_R$ would be flipped based on them. Figure 3 shows a quantum circuit that would achieve this example encoding. Note that the figure is in MSB format to align with the table, which is opposite of the traditional LSB format used in conventional quantum circuit diagrams.

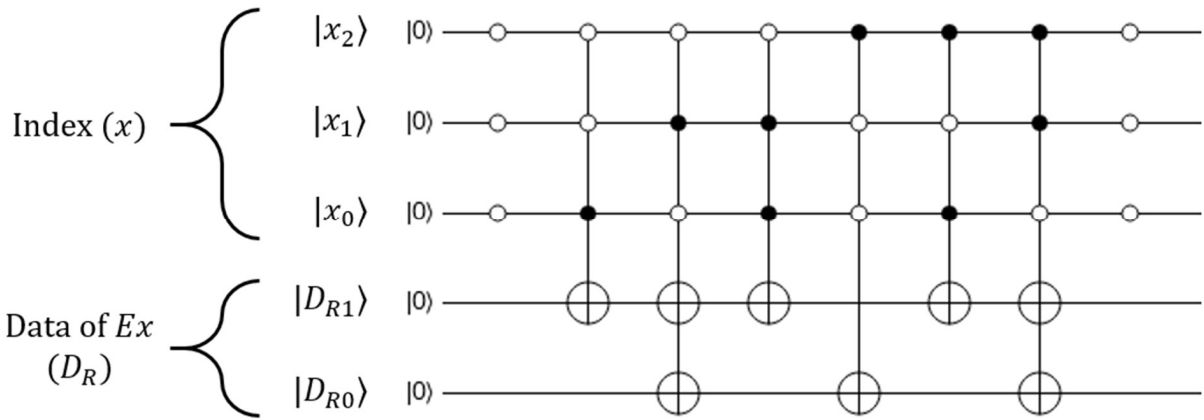

**Figure 3. Quantum circuit for the logic in Table 1.**

As demonstrated, this approach clearly does not scale practically and destroys any potential speedup the quantum computer could provide. To combat this, NEQR recommends employing the Espresso heuristic logic minimizer[3] to compress the circuit. Espresso has been used in the classical PLA field for over thirty years with great success, and open source implementations are available[4]. NEQR's Espresso scheme consists the following steps:

1. Create a logic table for the entire sequence similar to Table 1. The inputs will be the index bits. The outputs will be the value bits. Indices where the value is 0 can be ignored.

2. Run Espresso on this table using the "d1merge" command. This will create a compressed logic sequence for the input data that is compatible with quantum systems (i.e. each row can be treated as an independent set of controlled X gates).

3. Create a quantum circuit to perform the logic from Espresso's output.

---

[3] https://en.wikipedia.org/wiki/Espresso_heuristic_logic_minimizer
[4] https://github.com/jclapis/espresso-logic

Following this process, Espresso produces the circuit shown in Figure 4 when run against (11). Again, note that this diagram is in MSB format so it may appear inverted.

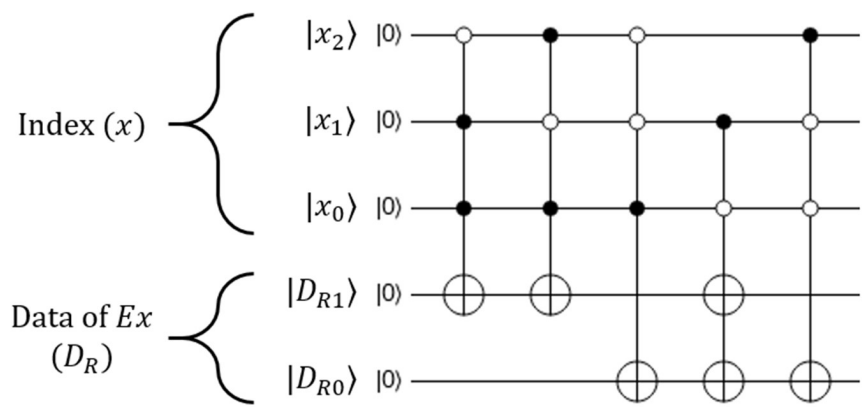

Figure 4. Quantum circuit for the Espresso output after being run on the example in (11).

This circuit is notably smaller than the brute-force circuit, requiring only 14 CCNOT gates (when using a CCNOT chain to achieve multicontrolled X gates, discussed in Appendix A5Appendix A) instead of the 24 from the circuit in Figure 3. A more thorough performance comparison between the brute-force and Espresso encoding schemes is included in Appendix B.

QDP uses the Espresso encoding system twice in this step: once to encode $S_R$ into $|x, D_R\rangle$ and once to encode $S_Q$ into $|y, D_Q\rangle$. After this step, the states in (9) and (10) have been achieved.

### 2.2.1 Padding the Sequences

If the sequences are not lengths that are a power of two, they can be padded in this step by appending data to them that will not interfere with the dot plot. To pad $S_R$, use a value that does not appear anywhere in $S_R$ or $S_Q$ called $P_R$. To pad $S_Q$, use a separate value that does not appear in either sequence and is different from $P_R$, called $P_Q$.

Continuing the DNA example, both sequences will be made of A, C, G, and T elements (assigned values of 0, 1, 2, and 3). To pad $S_R$, we would append the new value $P_R$ (which is assigned a value of 4) as many times as necessary to make $W = 2^w$. Similarly, to pad $S_Q$, we would append the new value $P_Q$ (which is assigned a value of 5) enough times to make $H = 2^h$. Because these values do not appear in either original sequence, and they will never match with their opposing sequences, they will not impact the dot plot in any way.

Note that in this example, the size of the data registers $d$ will need to increase from 2 to 3 in order to account for these padding values.

## 2.3 Quantum XOR

In this step, the actual dot plot image is generated. Recall from (5) that a dot plot is a 1-bit-per-pixel image where the pixel at $(x, y)$ is 0 (white) if $S_{R,x} \neq S_{Q,y}$ and 1 (black) if $S_{R,x} = S_{Q,y}$. This image would be generated classically by comparing each element to each other element, thus prompting $2^w * 2^h$ operations. Using quantum parallelism, we can achieve this with just one quantum XOR – an exponential speedup in its own right, even before considering the QFT at the end of QPSA.

We leverage an in-place quantum XOR ($\oplus$) operation on $D_R$ and $D_Q$ that achieves the following state:

$$|D_R, D_Q\rangle \rightarrow |D_R, D_R \oplus D_Q\rangle \tag{12}$$

Here $D_Q$ is transformed into the XOR of $D_R$ and $D_Q$. To achieve this, we simply apply CNOT on each pair of qubits where $D_{R,i}$ is the control and $D_{Q,i}$ is the target. Table 2 shows the state of the registers before and after this operation, thus confirming that it has the intended effect of achieving $D_R \oplus D_Q$.

| $D_{R,i}$ | $D_{Q,i}$ Before CNOT | $D_{Q,i}$ After CNOT | $D_{R,i} \oplus D_{Q,i}$ |
|---|---|---|---|
| 0 | 0 | 0 | 0 |
| 0 | 1 | 1 | 1 |
| 1 | 0 | 1 | 1 |
| 1 | 1 | 0 | 0 |

**Table 2. Demonstration of CNOT as an in-place XOR operation.**

Because $D_R$ and $D_Q$ are both in superposition of all possible values for their respective sequences (i.e. they represent all indices $x$ and $y$ simultaneously), applying CNOT in this fashion will perform an XOR of all values of these two registers simultaneously as well.

After this, $D_Q$ no longer represents $S_Q$, but instead captures the bitwise comparison of $D_R$ and $D_Q$ for all $x$ and $y$. More formally, these registers are now in this state:

$$|y, x, D_Q\rangle = \frac{1}{\sqrt{2^{h+w}}} \sum_{i=0}^{2^h-1} \sum_{j=0}^{2^w-1} |i, j, S_{R,i} \oplus S_{Q,j}\rangle \tag{13}$$

By the definition of the XOR operation, if $S_{R,i} = S_{Q,j}$, then $D_Q$ will be 0. If $S_{R,i} \neq S_{Q,j}$, then $D_Q$ will be any other value but 0. Thus, we can use a multi-zero-controlled X gate, using $D_Q$ as the controls and the single qubit $v$ as the target, to create the dot plot. Figure 5 shows a circuit diagram of this step.

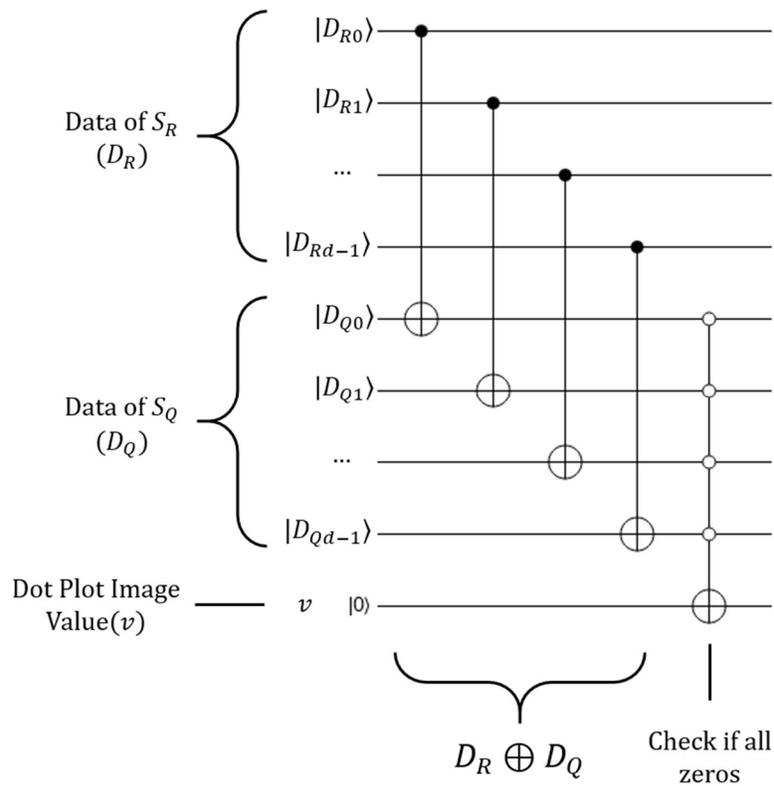

**Figure 5. Circuit for step 3 of QDP (dot plot generation).**

After this step, the registers $y$, $x$, and $D_Q$ now represent the required state for the QPSA Black Box, and QPR can be run normally. Figure 6 shows a complete circuit diagram of QPR using QDP as the Black Box.

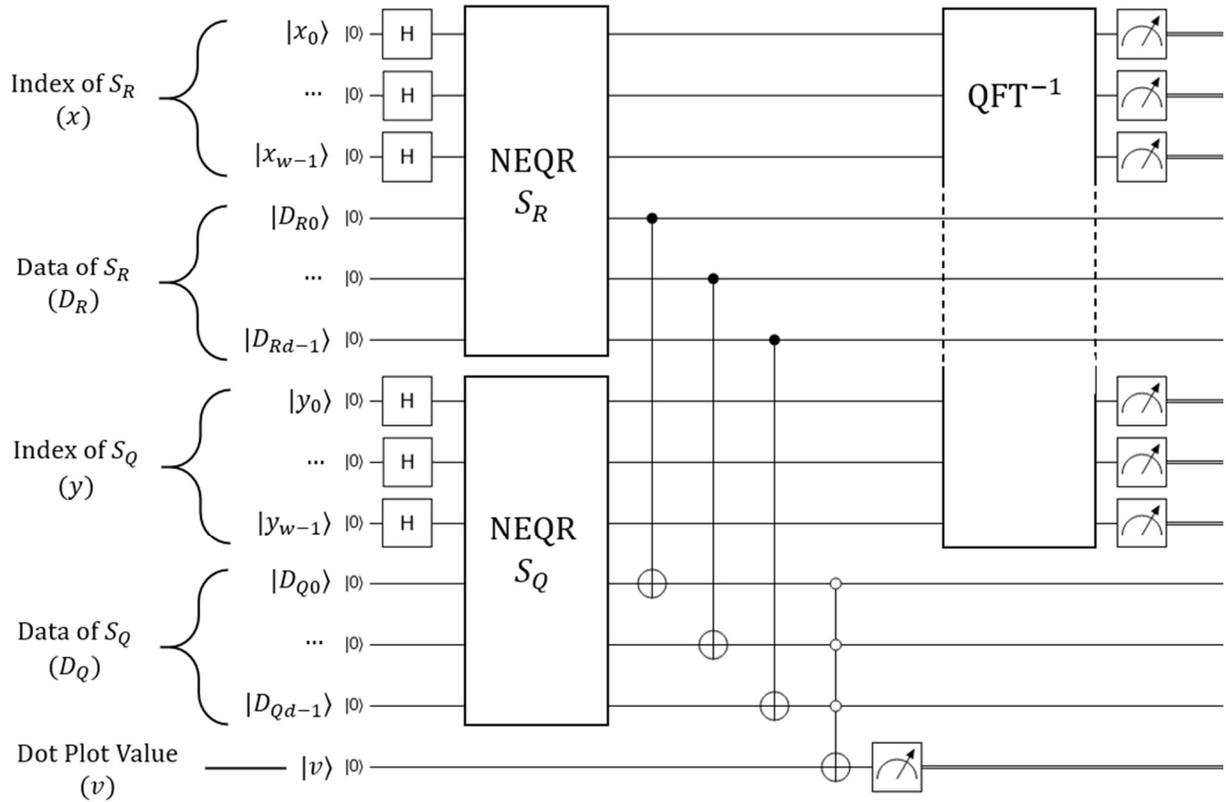

**Figure 6. Complete circuit diagram of QPR, using QDP as the Black Box.**

Note that the QFT is run on a combination of the $x$ and $y$ registers; despite its visual placement in the diagram, the data register $D_R$ is not included in it, as indicated by the dashed line.

# 3 Runtime Analysis and Performance Measurements

We assessed QDP with a combination of mathematical analysis techniques and empirical data produced by implementing QPR with QDP in software using the QDK v0.10.2002.2610[5] and Qiskit v0.15.1[6] frameworks.

## 3.1 General Complexity Analysis

Mathematically, the total qubit count required for QDP is $x + y + 2d + 1$, plus an additional set of ancilla qubits required to decompose multicontrolled X gates into natively supported gates. The exact decomposition method used (discussed in Appendix A) will impact the width. With the basic CCNOT chain method, it will require up to $x - 2$ or $y - 2$ ancilla qubits, whichever is larger. For sequences that have the same bit length $n$, QPR with QDP will have the following width requirement:

$$2n + 2d + 1 \leq \text{Width} \leq 3n + 2d - 1 \qquad (14)$$

---

[5] https://www.microsoft.com/en-us/quantum/development-kit
[6] https://qiskit.org/

The exact number depends on the sequences, and thus on the control logic generated by Espresso in QDP's step 2. In either case, this represents an $O(n)$ circuit width. The data size $d$ is effectively constant for a given domain (such as nucleotide sequences), so its contribution to the circuit width does not increase as the sequences themselves grow.

Continuing the DNA example from section 2, this would take between 21 and 27 qubits which is still an exponential decrease over the 65,536 bits required to create the classical dot plot.

The depth of the circuit is best described in parts. The sequence encoding step using NEQR's Espresso encoder, while compressed, is expected to scale exponentially in the bit size of the sequence lengths. The dot plot generation step does not depend on the sequence length, but rather on the bit size of the data registers. It requires one CNOT for every additional qubit in the registers, so this step will scale linearly in the bit size of the registers giving a complexity of $O(d)$. Since this size is fixed and relatively small (compared to the sequence sizes) for most conventional use cases, we anticipate that this will be the smallest contributor to the overall depth – even smaller than the QFT (of complexity $O(n)$) which is required at the end of QPR. The overall depth is thus dominated by the sequence encoding scheme, which in its current form produces an $O(2^n)$ circuit depth.

## 3.2 Software Simulation Results

We used three sets of data in our software simulations, which were retrieved from the National Center for Biotechnology Information's nucleotide database[7]:

- Dataset 1 compared two DNA sequences[8], both of which were 256 elements (base pairs) long.
- Dataset 2 compared one sequence[9] of length 1,024 base pairs against itself.
- Dataset 3 compared two sequences[10] of length 4,096 base pairs.

We selected these sequences at random from the database simply because their lengths were a power of two, which meant that we didn't have to pad them. In all of these cases, the sequences contained four possible elements (A, C, G, or T), so we used $d = 2$ for the size of our data registers.

### 3.2.1 Code Validation

We tested our implementation of the NEQR Espresso-based encoder with two methods. The following breakdown shows the process for method 1:

1. Prepare $k$ as the index register and $v$ as the value register. Rather than using the Hadamard gates in step 1 of QDP, explicitly set $k$ to one specific index using the appropriately placed X gates. Note that it will not be in superposition.
2. Generate the circuit for NEQR Espresso-based encoding against $k$ and $v$.

---

[7] https://www.ncbi.nlm.nih.gov/nuccore
[8] https://www.ncbi.nlm.nih.gov/nuccore/AJ301403.1 vs. https://www.ncbi.nlm.nih.gov/nuccore/AJ301440.1
[9] https://www.ncbi.nlm.nih.gov/nuccore/12711313
[10] https://www.ncbi.nlm.nih.gov/nuccore/NZ_QOXC01000078.1 vs. https://www.ncbi.nlm.nih.gov/nuccore/NZ_NLIT01000082.1

3. Append measurements to the circuit on all qubits of $k$ and $v$, measuring them both as integers.
4. Simulate this circuit using QDK's Toffoli simulator[11].
5. Compare the measured values of $k$ and $v$ with the original sequence as a reference. Ensure that the index selected in step 1 was measured in $k$, and that its corresponding value $v$ is correct.
6. Repeat steps 1-4 for all possible indices of the sequence, thus ensuring that the algorithm correctly encodes the sequence for every element.

The following list shows the process for method 2:

1. Prepare $k$ normally, using the Hadamard gate to put it into a complete uniform superposition.
2. Generate the circuit for NEQR Espresso-based encoding as in method 1.
3. Append the measurements on $k$ and $v$ as in step 2.
4. Simulate the circuit using Qiskit's Aer simulator[12], indicating a large number of shots (e.g. 100,000).
5. Compare each resulting set of measurements against the original sequence. The distribution of measured $k$ values should be random and uniform, and each accompanying $v$ measurement should be correct.

Method 1 exhaustively proves that the implementation works for every value in the provided sequence, and method 2 confirms its correctness while operating on qubits in superposition.

We validated our implementation of the complete QDP algorithm by using both of these methods, but this time we compared the resulting measured values of $x$, $y$, and $v$ against a classically generated dot plot image as a reference.

To assess QPR as a whole, we used the implementation of the inverse QFT provided by QDK and Qiskit. We validated these implementations separately in prior work and did not include them in our software validation here.

### 3.2.2 Quantum Platforms and Assembly Generation

The produced circuits for datasets 2 and 3 are fairly elaborate and require substantial classical computational resources to simulate. Luckily, for the purposes of runtime analysis and resource estimation, we do not need to actually simulate the circuits. Instead, we only need to explore the quantum machine code that will execute on a quantum computer.

---

[11] The Toffoli simulator is a similar to a full statevector simulator, but limits the valid gates to X and controlled X with arbitrarily many control qubits (e.g. CNOT, CCNOT, 5-CNOT). Because of this, it is extremely fast and efficient for what are effectively classical binary logic circuits, but it does not support superpositions. For more information, consult the documentation at https://docs.microsoft.com/en-us/quantum/machines/toffoli-simulator.

[12] The Aer simulator is a full statevector simulator with no restrictions. It simulates measurements by first calculating the final statevector for the circuit, then randomly sampling that statevector as many times as requested (rather than sampling it once and re-simulating the entire circuit for each requested run). To enable this, measurements are restricted to the end of a circuit. For more information, consult the documentation at https://qiskit.org/aer/.

For our software-based empirical measurements, we elected to study the OpenQASM [8] assembly generated by Qiskit's transpiler. QDK provides a resource estimator but it operates on the circuits in a vacuum; it does not consider the details of a given hardware backend such as qubit connection topology and native gate set, all of which are crucially important when constructing platform-specific quantum assembly code. Furthermore, it does not currently produce quantum assembly code, which means it cannot be used to execute circuits on physical machines. While we included QDK's resource estimates for comparison, we exclusively leveraged Qiskit's transpiler to explore the runtime requirements of QDP and QPR on real quantum computers.

We selected three target platforms for evaluation:

- The Aer simulator, which natively supports many gates including H, X, $U_1$, $U_2$, $U_3$, CNOT, CCNOT, $CU_3$, and SWAP (all of which were present in our compiled assembly).

- IBM Q's Rochester architecture[13], which has 53 superconducting qubits and natively supports the $U_1$, $U_2$, $U_3$, and CNOT gates. As with all superconducting architectures, it has limited qubit connectivity; its coupling map is described in the source code of the mock backend we used for testing[14].

- Alpine Quantum Technologies's Innsbruck architecture[15], which has 4 trapped-ion qubits and natively supports the $R_X$, $R_Y$, and $R_{XX}$ gates. For the purposes of our tests, we arbitrarily inflated the number of supported qubits to 40; because this is a trapped ion system, it allows unrestricted any-to-any qubit connectivity and thus increasing the qubit count will not impact the measured performance statistics of the generated assembly code.

### 3.2.3 Resource Estimations

We ran Qiskit's transpiler to generate circuits for datasets 1 through 3 with several different parameters. For each one, we collected the following measurements:

- Espresso compression time
- Number of multicontrolled X gates in Espresso-based control set
- Multicontrolled X decomposition mode (see Appendix A)
- Physical backend system being compiled for
- Time taken to construct the initial circuit as a Qiskit circuit object
- Time taken to transpile the circuit into backend-specific QASM code
- Total circuit width (number of qubits required to execute it)
- Circuit depth (number of gates in the critical, or longest serial, path) for the NEQR encoding step

---

[13] https://newsroom.ibm.com/2019-09-18-IBM-Opens-Quantum-Computation-Center-in-New-York-Brings-Worlds-Largest-Fleet-of-Quantum-Computing-Systems-Online-Unveils-New-53-Qubit-Quantum-System-for-Broad-Use
[14] https://github.com/Qiskit/qiskit-terra/blob/master/qiskit/test/mock/backends/rochester/fake_rochester.py
[15] https://www.aqt.eu/technology/

- Circuit depth for the dot plot generation step
- Circuit depth for the QFT at the end of QPR (included for reference)
- Total circuit depth for the complete QPR circuit, combining all of the previous steps; note that this is not a simple addition of the depths in each step, because when combined into a single circuit, the optimizer is able to reduce the total count slightly
- Raw counts for each individual gate type (e.g. total number of H and CNOT gates; varies for each backend architecture based on their supported native gate sets)

Qiskit's transpiler offers 4 optimization levels, labeled 0 (None), 1 (Light), 2 (Medium), and 3 (Heavy). We selected to use optimization mode 1 (Light), which offered the best balance of compilation time and circuit depth.

We have produced subsets of the captured measurements in the following tables, one per dataset. The raw gate counts are not included, but are available upon request.

For reference, our test machine had two Xeon E5-2687W v4 CPUs at 3.00 GHz (24 total cores), and 220 GB of DDR4 RAM at 2667 MHz. The core count is not relevant because Qiskit's transpiler is single-threaded; the RAM ended up being the main contributing factor with respect to overall performance, with many of the larger trials requiring over 100 GB.

| SDK | MCX Mode | Backend | Circuit Creation Time | Compile Time | Width | NEQR Depth | QDP Depth | QFT Depth | Total Depth |
|---|---|---|---|---|---|---|---|---|---|
| QDK | --- | Resource Counter | 00:00.8 | --- | 27 | 5925 | 5 | 114 | 6026 |
| Qiskit | CCNOT Chain | Aer Sim | 00:01.3 | 00:08.3 | 27 | 14,672 | 4 | 33 | 14,698 |
| Qiskit | Single-Qubit | Aer Sim | 00:03.7 | 00:19.9 | 22 | 48,803 | 4 | 33 | 48,831 |
| Qiskit | CCNOT Chain | Rochester | 00:01.2 | 07:27.9 | 53 | 38,196 | 21 | 590 | 38,722 |
| Qiskit | Single-Qubit | Rochester | 00:03.7 | 34:49.6 | 53 | 154,146 | 21 | 590 | 154,734 |
| Qiskit | CCNOT Chain | AQT Innsbruck | 00:01.1 | 00:46.7 | 27 | 35,039 | 34 | 330 | 35,307 |
| Qiskit | Single-Qubit | AQT Innsbruck | 00:03.7 | 03:57.1 | 22 | 238,358 | 34 | 330 | 238,652 |

Table 3. Resource estimates for dataset 1 (two 256-element sequences).

| SDK | MCX Mode | Backend | Circuit Creation Time | Compile Time | Width | NEQR Depth | QDP Depth | QFT Depth | Total Depth |
|---|---|---|---|---|---|---|---|---|---|
| QDK | --- | Resource Counter | 00:00.1 | --- | 33 | 31,301 | 5 | 146 | 31,447 |
| Qiskit | CCNOT Chain | Aer Sim | 00:06.3 | 00:44.9 | 33 | 80,918 | 4 | 41 | 80,950 |
| Qiskit | Single-Qubit | Aer Sim | 00:20.8 | 01:53.7 | 26 | 260,801 | 4 | 41 | 260,835 |

| SDK | MCX Mode | Backend | Circuit Creation Time | Compile Time | Width | NEQR Depth | QDP Depth | QFT Depth | Total Depth |
|---|---|---|---|---|---|---|---|---|---|
| Qiskit | CCNOT Chain | Rochester | 00:06.3 | 38:12.0 | 53 | 213,473 | 20 | 839 | 214,234 |
| Qiskit | Single-Qubit | Rochester | 00:32.2 | 4:07:21.9 | 53 | 833,199 | 20 | 839 | 833,995 |
| Qiskit | CCNOT Chain | AQT Innsbruck | 00:06.4 | 04:19.5 | 33 | 173,664 | 34 | 418 | 174,000 |
| Qiskit | Single-Qubit | AQT Innsbruck | 00:20.6 | 21:55.6 | 26 | 1,272,499 | 34 | 418 | 1,272,870 |

Table 4. Resource estimates for dataset 2 (one 1024-element sequence against itself).

| SDK | MCX Mode | Backend | Circuit Creation Time | Compile Time | Width | NEQR Depth | QDP Depth | QFT Depth | Total Depth |
|---|---|---|---|---|---|---|---|---|---|
| QDK | --- | Resource Counter | 00:01.8 | --- | 39 | 150,239 | 5 | 178 | 150,417 |
| Qiskit | CCNOT Chain | Aer Sim | 00:45.2 | 05:07.3 | 39 | 384,312 | 4 | 49 | 384,350 |
| Qiskit | Single-Qubit | Aer Sim | 03:54.6 | 18:38.7 | 30 | 1,940,132 | 4 | 49 | 1,940,173 |
| Qiskit | CCNOT Chain | Rochester | 00:53.9 | 4:34:26.7 | 53 | 1,027,772 | 20 | 1108 | 1,028,758 |
| Qiskit | Single-Qubit | Rochester | DNF[16] | | | | | | |
| Qiskit | CCNOT Chain | AQT Innsbruck | 00:43.8 | 30:20.3 | 39 | 761,606 | 34 | 506 | 762,002 |
| Qiskit | Single-Qubit | AQT Innsbruck | 02:53.0 | 2:50:15.7 | 30 | 9,524,360 | 34 | 506 | 9,524,819 |

Table 5. Resource estimates for dataset 3 (two 4096-element sequences).

Our results indicate several trends which describe QDP's resource requirements.

From a circuit width perspective, the unrestricted topology systems (the Aer simulator and the AQT trapped ion machine) met the theoretical predictions (21 to 27 for dataset 1, 25 to 33 for dataset 2, 29 to 39 for dataset 3, all depending on the MCX mode used). For the superconducting Rochester architecture, the Qiskit transpiler elected to use all 53 available qubits in for all three datasets. This demonstrates that while our predictions were accurate for any-to-any topology systems, they serve as lower bounds on nearest-neighbor topology systems; the transpiler may simply consume all available qubits to minimize the number of SWAP gates in the circuit. The exact lower bound on the required width depends on the machine's coupling map and is difficult to predict mathematically, but is revealed through algorithm compilation.

Note that in our exploration, we assume ideal qubits – that is, error correction is not required. In reality, error correction is mandatory for execution on NISQ devices such as those we have compiled against here. Applying this would necessarily inflate the circuit width by orders of magnitude, depending upon the fidelity of the underlying gates, but for the purpose of this analysis, we can safely disregard it.

---

[16] This test consumed all of the available system RAM (220 GB), ran for several hours, and ultimately crashed with an out-of-memory (heap allocation) error.

The use of the single-ancilla multicontrolled X mode had a notable impact on the width for any-to-any topology systems when compared to the CCNOT chain mode, bringing it from the upper bound to the lower bound (plus one). However, this increased the depth by an order of magnitude in every trial. This cost is almost never worth the reduction in width, especially when QDP requires relatively few qubits in the first place. Our findings suggest that the basic CCNOT chain is the preferred choice in any situation where the target system has enough qubits to execute it.

We observe that the NEQR circuit depth is roughly five times larger between each dataset. This trend is consistent across all three backend systems. To confirm the trend as exponential growth, we executed NEQR on four additional sequences of lengths 512, 2048, 8192, and 16384 base pairs (9, 11, 13, and 14 bits, respectively)[17] against themselves. These were limited to the Aer simulator backend, using the CCNOT chain mode, because this configuration offered the fastest compilation time but was still consistent with the physical hardware backend results in the above trials. The results are shown in Table 6 and Figure 7.

| Sequence Length (Bits) | Total Run Time (Seconds) | Circuit Depth | $\log_2 Depth$ |
|---|---|---|---|
| 8 | 9.604 | 14,672 | 13.84 |
| 9 | 21.876 | 34,178 | 15.06 |
| 10 | 51.591 | 80,918 | 16.30 |
| 11 | 102.275 | 180,980 | 17.47 |
| 12 | 365.854 | 384,312 | 18.55 |
| 13 | 615.295 | 867,454 | 19.73 |
| 14 | 1272.156 | 1,857,560 | 20.82 |

Table 6. QDP circuit depths for sequence lengths of 8 to 14 bits (Aer simulator backend, CCNOT chain mode).

---

[17] 512 base pair sequence: https://www.ncbi.nlm.nih.gov/nuccore/MT000611.1
2048 base pair sequence: https://www.ncbi.nlm.nih.gov/nuccore/NZ_WOTD01000116.1?report=gbwithparts
8192 base pair sequence: https://www.ncbi.nlm.nih.gov/nuccore/NZ_JAAACW010000051.1?report=gbwithparts
16384 base pair sequence: https://www.ncbi.nlm.nih.gov/nuccore/NZ_SLXT01000041.1?report=gbwithparts

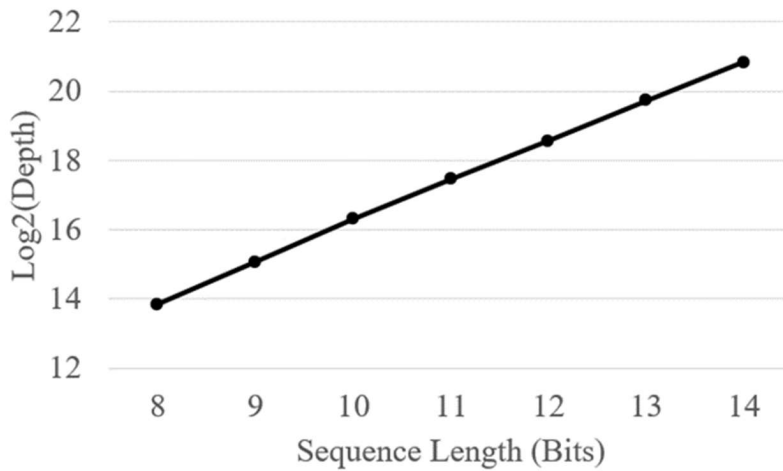

**Figure 7. Bit size of sequence length vs. $\log_2 Depth$ for QDP (Aer simulator backend, CCNOT chain mode).**

These results confirm that the circuit's depth grows exponentially with the number of bits in the sequence's length.

As expected, the dot plot generation step only provides a small, constant contribution to the overall depth regardless of sequence size because the size of the data registers does not change. The QFT step scales linearly in the bit size of the sequence. Both of these steps are insignificant compared to the NEQR encoding step.

### 3.2.4 Runtime Estimation

In [1], the authors compare the runtime performance of QPSA against four classical pairwise sequence aligners. The analysis aligned *E. coli K12 MG1655*[18] with *E. coli 536*[19]. These sequences are very large; the former contains 749.9M base pairs, and the latter contains 4.94M base pairs. To accommodate for such sizes, the authors segmented the first sequence into a collection of 20.8 million separate sets of 200 base pairs each. The second sequence was divided into segments of 1000 base pairs each; though it is not specified in the paper, we assume they used 4940 segments which would account for the entire sequence. Each of the 20.8M segments of sequence 1 were then aligned with each segment of sequence 2.

The authors reported that according to their estimates, QPSA was able to align these sequences in 28.84 seconds, demonstrating 94.24% (± 1.48) precision and 97.83% (± 0.85) recall. By comparison, the fastest classical algorithm, CUSHAW [9], aligned them in 3711 seconds with 90.00% precision and 97.51% recall. In this test, QPSA was two orders of magnitude faster than the classical algorithm. However, the authors state in their analysis that they assume the Black Box has the same runtime as QFT. Our analysis has shown that using QDP for the Black Box does not satisfy this assumption, as QDP scales exponentially in the sequence length's number of bits, whereas QFT scales linearly.

---

[18] https://www.ncbi.nlm.nih.gov/sra/SRR001665
[19] https://www.ncbi.nlm.nih.gov/nuccore/NC_008253.1/

We explored QDP's runtime characteristics using a variant of this setup. We used QDP to align the first sequence from dataset 1 in our original trials (256 base pairs) with the sequence from dataset 2 (1024 base pairs). This roughly approximates a single execution of QPSA's alignment subroutine in the 200-vs-1000 configuration used by the authors in their original analysis. We used the CCNOT chain mode as it offered the best runtime. The results are presented in Table 7.

| Backend | Circuit Time | Transpile Time | Depth | Estimated Execution Time (sec) |
|---|---|---|---|---|
| **Aer Sim** | 00:04.79 | 00:30.12 | 48,087 | --- |
| **Rochester** | 00:04.66 | 24:36.67 | 127,315 | 0.0166 |
| **AQT Innsbruck** | 00:03.93 | 02:43.13 | 105,143 | 2.1029 |

Table 7. Runtime estimates for a single iteration of QPSA using QDP as the Black Box when aligning a sequence of length 256 against one of length 1024.

To determine the total execution time on each backend system, we used the data presented in [10]: we assigned IBM's Rochester system a time of 130 ns per gate, and AQT's Innsbruck system a time of 20 μs per gate. Note that these are the times of single-qubit gates only; we did not break the circuit's critical path down into single and multi-qubit gates so these estimates are quite conservative.

To reproduce the same alignment test performed in [1] and [9], a circuit of comparable length would need to run millions of times in order to cover the entirety of the two *E. coli* sequences, which would bring the total execution time well into the realm of years, rather than seconds. This does not take the overhead associated with circuit compilation into account, which itself is several orders of magnitude larger than the actual circuit execution times.

To reiterate, this analysis ignores the invariable requirement of error correction for execution on NISQ devices. Such functionality would further inflate the execution time and circuit width, but QDP has already been shown to be impractical for conventional sequence lengths even without it.

# 4 Conclusions

In this report we have presented an algorithm, the quantum dot plot generator (QDP), that can implement the unknown Black Box required by the quantum pairwise sequence aligner (QPSA). The algorithm leverages quantum parallelism to quickly and efficiently generate the 2D dot plot of two arbitrary sequences (a map of the equality of the elements of those sequences for all pairs of indices). We have demonstrated that quantum circuits for this algorithm can be reliably constructed and executed on universal gate-based quantum computers without requiring any special-purpose hardware.

QPSA with QDP is broken into three stages: encoding the original sequences in NEQR format, generating the dot plot image, and performing the QFT. We estimated the runtime of each step by implementing QPSA and QDP in software with the Qiskit and QDK frameworks, and using their built-in resource counters to analyze the resulting compiled quantum programs. Our analysis demonstrated the following key points:

- Overall circuit widths are exponentially reduced compared to classical memory requirements during all steps of the process. This means large sequences of data can be efficiently encoded and processed in a comparatively small number of qubits. Aligning typical sequences with $2^{30}$ elements would typically require 93 fault-tolerant qubits using the fastest multicontrolled X implementation, though systems with nearest-neighbor interaction topologies may require a few more.

- Sequence encoding requires a linear number of gates with respect to the sequence length (between 761,606 and 1,027,772 gates for 12-bit sequence lengths, depending on the target architecture), which defeats any exponential computational speedup.

- Dot plot generation requires a logarithmic number of gates with respect to the data size. It is not affected by sequence length. For most use cases, it comes with a very small fixed overhead cost (20 to 34 gates for 2-bit data, depending on the target architecture).

- The QFT requires a logarithmic number of gates with respect to the sequence length. However, decomposition into the native gate set supported by the target architecture does give it a non-trivial cost (506 to 1108 gates for 12-bit sequence lengths, depending on the target architecture) which will impact runtime compared to theoretical estimates.

Because of the lack of exponential speedup in the encoding step, the algorithm as a whole does not offer any practical advantage over classical computation; QDP is relegated to the ever-increasing list of promising quantum algorithms that fall victim to this particular shortcoming. We suggest that future work in this domain focus on systems for encoding classical data in quantum registers while achieving an exponential gate reduction, such as a reliable quantum RAM device [11]. Such a system would bring QDP's execution requirements down to the realm of practicality, and thus enable researchers to use QPSA on universal quantum computers (which we anticipate may be coupled to qRAM systems for general-purpose usage in the future). Until such a method of efficiently encoding large amounts of classical data into qubits is achieved, QDP (and by extension, QPSA) will not deliver an exponential speedup on universal gate machines, regardless of platform maturity.

# Appendix A – Decomposition of Multicontrolled X Gates

Universal quantum computers today do not inherently provide support for controlled X gates that have more than one control qubit (e.g. CCNOT and beyond), which are key in the NEQR sequence encoding and the check-if-all-zeroes steps of QDP. There are several ways to synthesize these gates using the more primitive, readily-available gates from the computer's native gate set. We compared two such methods during our analysis.

## The Basic CCNOT Chain

In this method, controlled X gates with arbitrary many control qubits are decomposed into a succession of CCNOT gates as shown in Figure 8.

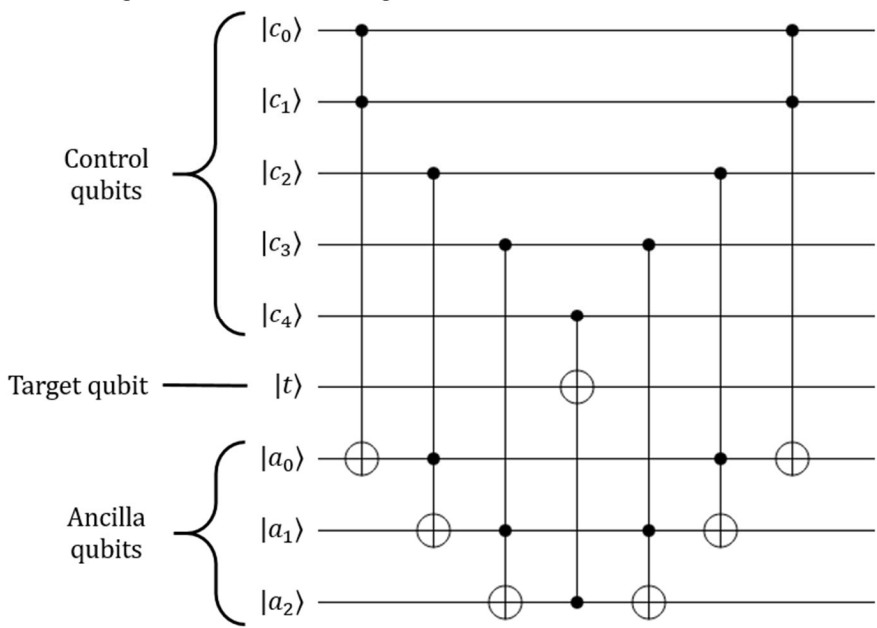

**Figure 8. Decomposing a multicontrolled X gate via the CCNOT chain method.**

This method requires the use of ancilla (spare) qubits to temporarily "cache" the results of each intermediate CCNOT. The total number of ancilla required is $c - 2$, where $c$ is the number of control qubits in the operation. This is a relatively fast and efficient method of achieving multicontrolled X gates, though the specifics depend on how the provided quantum computer deconstructs CCNOT into its native gate set.

## The Advanced Single-Ancilla Technique

This method uses more advanced multicontrol techniques presented in [12] which only require a single ancilla qubit. Qiskit opts to use these techniques to implement primitive 3-CNOT and 4-CNOT gates (X gates controlled with 3 and 4 qubits, respectively). Arbitrarily-controlled X gates are then recursively decomposed into a set of CNOT, CCNOT, 3-CNOT, and 4-CNOT gates appropriately. Since Qiskit is open source, one can review their implementation details if interested[20] but they are reproduced here for convenience.

---

[20] https://github.com/Qiskit/qiskit-terra/blob/master/qiskit/extensions/standard/multi_control_toffoli_gate.py

The 3-CNOT gate is constructed as shown in Figure 9:

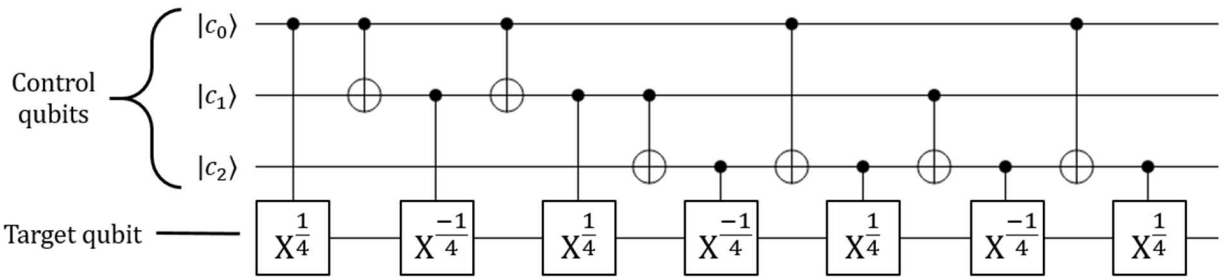

**Figure 9. Circuit diagram for the 3-CNOT gate.**

The 4-CNOT gate is constructed as shown in Figure 10:

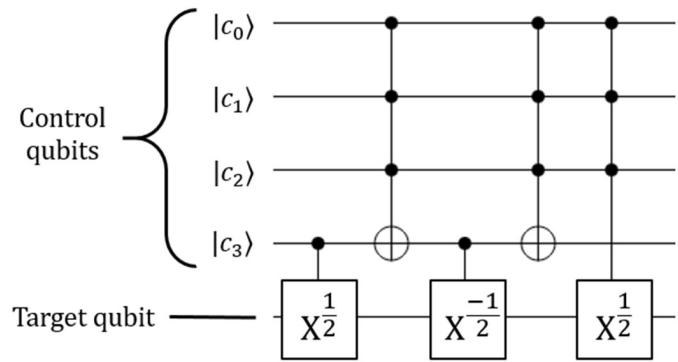

**Figure 10. Circuit diagram for the 4-CNOT gate.**

The final gate of this circuit ($\sqrt{X}$ controlled by 3 qubits) is achieved by the 3-CNOT circuit in Figure 9, but using the gates $X^{1/8}$ and $X^{-1/8}$ in place of the $X^{1/4}$ and $X^{-1/4}$ gates, respectively. The single ancilla qubit is used as a cache between recursive decompositions into 1- to 4-CNOT gates. This technique has a width advantage, as it costs fewer ancilla qubits, but it comes with a significant increase in depth (nearly an order of magnitude over the basic CCNOT chain, as shown in section 3.2.3).

# Appendix B – Espresso Efficacy

Prior to running the complete test suite described in section 3.2.3, we experimented with the NEQR encoder to observe the differences in circuit length between the "brute-force" approach and the Espresso approach. We implemented both schemes in QDK and used its resource counter to quickly build and analyze the resulting circuits. A comparison of running this test on the three datasets from the main report is shown in Table 8.

| Dataset | Espresso Runtime (Seconds) | MCX Count (Brute Force) | MCX Count (Espresso) | QDP Depth (Brute Force) | QDP Depth (Espresso) | Espresso Compression (%) |
|---|---|---|---|---|---|---|
| 1 | 0.271 | 379 | 232 | 22420 | 5925 | 73.57% |
| 2 | 0.287 | 1476 | 928 | 113652 | 31301 | 72.46% |
| 3 | 0.288 | 5636 | 3446 | 544540 | 150239 | 72.41% |

**Table 8. Comparison of brute-force and Espresso-based NEQR encoding schemes.**

In all three of our datasets, Espresso reduced the number of multicontrolled X gates by an average of 38.26%. Normally one would assume this directly translates to the resulting circuit depth reduction, but some of the Espresso control sets have fewer control qubits, which further reduces the total circuit depth. We saw a consistent reduction of 72% to 74% in circuit depth across all three data sets. Given the fact that Espresso runs in less than 1 second for the sizes of the sequences we used in our datasets and produces such impressive compression ratios, it appears to be a good compression candidate for any use case where classical data needs to be encoded in an NEQR-style superposition and entanglement scheme and the exponential depth growth of the circuits it produces is not a concern.